**ReadsMap: a new tool for high precision mapping of DNAseq and RNAseq read sequences**

Igor Seledtsov[1], Jaroslav Efremov[1,2], Vladimir Molodtsov[1,2], Victor Solovyev[1]

[1]Softberry Inc., Mount Kisco, NY, USA; [2]Novosibirsk State University, Novosibirsk, Russia

**Abstract**

There are currently plenty of programs available for mapping short sequences (reads) to a genome. Most of them, however, including such popular and actively developed programs as **Bowtie, BWA, TopHat** and many others, are based on Burrows-Wheeler Transform (BWT) algorithm. This approach is very effective for mapping high-homology reads, but runs into problems when mapping reads with high level of errors or SNP. Also it has problems with mapping RNASeq spliced reads (such as reads that aligning with gaps corresponding intron sequences), the kind that is essential for finding introns and alternative splicing gene isoforms. Meanwhile, finding intron positions is the most important task for determining the gene structure, and especially alternatively spliced variants of genes. In this paper, we propose a new algorithm that involves hashing reference genome. **ReadsMap** program, implementing such algorithm, demonstrate very high-accuracy mapping of large number of short reads to one or more genomic contigs. It is achieved mostly by better alignment of very short parts of reads separated by long introns with accounting information from mapping other reads containing the same intron inserted between bigger blocks.
**Availability and implementation: ReadsMap** is implemented in C. It is incorporated in **Fgenesh++** gene identification pipeline and is freely available to academic users at Softberry web server www.softberry.com.

Over the last years, development and improvement of next generation sequencing (NGS) methods has led to accumulation of enormous volumes of sequencing data, not only genomic, but also of transcripts. Vast majority of raw data consists of short reads [1], which need to be processed and analyzed, in order to reconstruct genomic sequences or those of the transcripts [2,3], or to determine gene expression profiles [4]. In order to perform such tasks, we need the instruments of mapping short reads to large sequences. Mapping refers to the process of aligning short reads to a reference genome sequence. While studying expression profiles doesn't require direct accounting for splice sites and intron positions, such steps are essential for predicting genes and alternative splice variants. Many of currently available read mapping programs effectively map unspliced reads [5,6], they aren't well suited for working with reads including two or more exons [7,8]. Most current programs, including such popular ones as Bowtie, BWA and TopHat,[7,8] are based on one of three widely used approaches: building a suffix tree, indexing reference genome using Burrows-Wheeler algorithm [9], or hashing reference genome. One of the programs (Pass) uses both hashing reference genome and exact dynamic programming alignment of a narrow region around the initial match [7,8]. Such approaches, however, either can't find splice sites at all, as is the case with Bowtie and BWA, or run into difficulties if such sites are located close to the ends of reads. To solve this problem, we proposed a novel approach implemented in ReadsMap program.

## Description of reads mapping method:

Mapping reads to chromosome sequences (or contigs) consists of the following principal steps (Fig.

1).

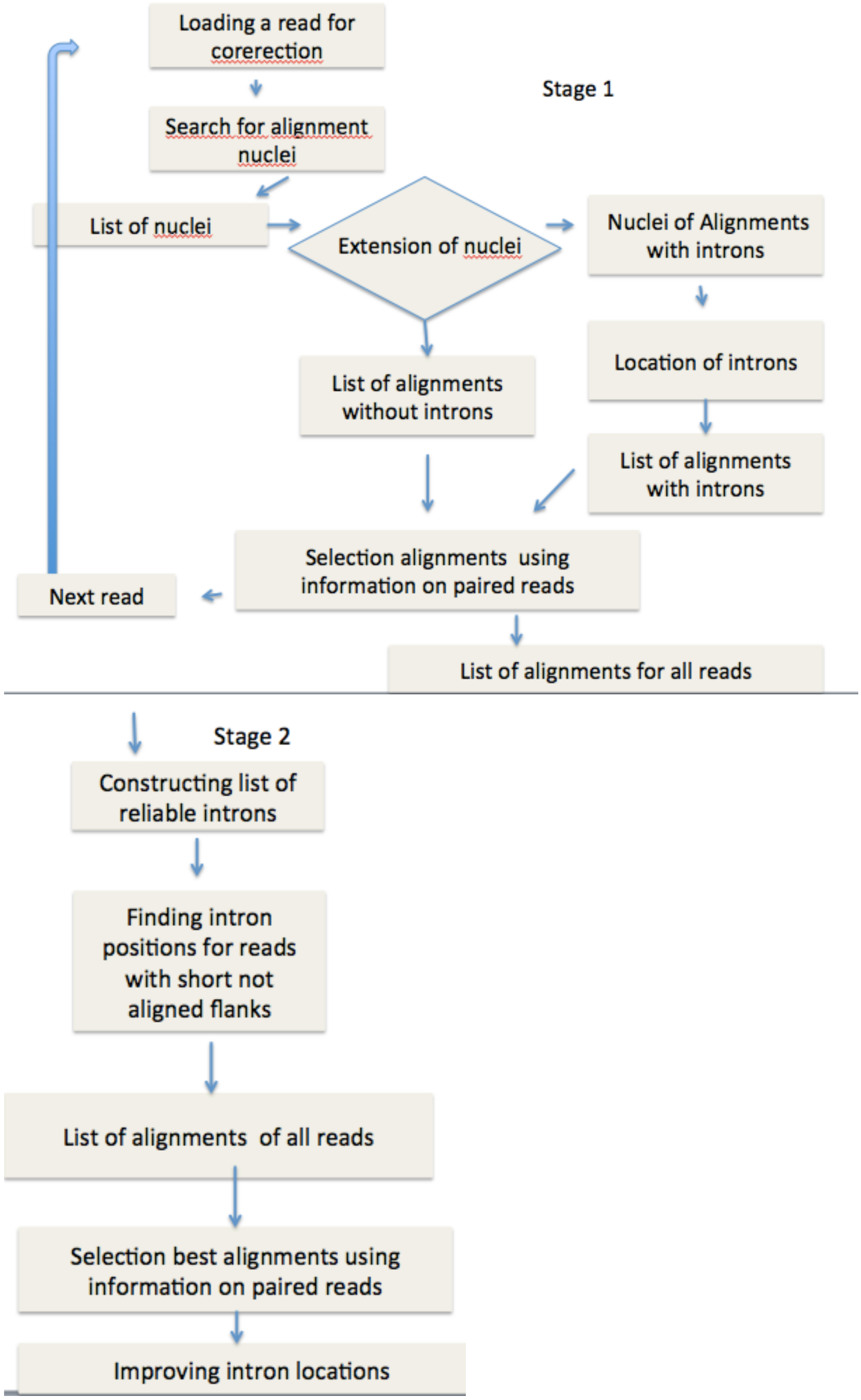

**Figure 1.** Work flow in ReadsMap program (Stage 1 and 2): loading a read for correction; removing low-quality terminal nucleotides; search for alignment nuclei; adding to a list of nuclei; extension of nuclei; adding to a list of spliced alignments; adding to a list of non-spliced alignments; selecting alignments supporting paired reads; loading the next read for analysis; generating a list of all reads' alignments; construction of list of reliable introns; localization of introns in alignments of reads with short unaligned flanks; building a list of all types of alignments; selection of best paired reads positions;

second iteration of computing intron position

**Alignment of a single read**

In the first step of a read alignment we search for so-called alignment nucleus, which is then used as a start point for constructing an entire alignment. In essence, nucleus is a region of high similarity between a read and a contig, which satisfies certain minimum requirements of length and similarity level. These requirements are calculated based on reads length, assumed number of introns, and a minimum length of exons. For instance, for reads of 50, 76 and 100 bp minimum lengths are 13, 20 and 26 bp correspondingly with homology level of 91.5%.

To minimize computational time required for finding such blocks, we scan only regions that exceeding predetermined number of matched k-mers (default k=12, and a minimum number of matches is 7). Such approach can be visualized as a rectangular matrix, one side of which corresponds to a reference sequence, and another – to a query read. Matches of k-mers of reference and query sequences (hits) are shown as dots grouped along diagonals of a matrix, while a particular k-mer can be observed in several diagonals. After building such matrix, we have a set of diagonals, each containing a set of hits. To select potential alignment nuclei we choose the diagonals with number of hits >=7.

Obviously, short segments of reads can almost always be aligned to multiple sites on a reference sequence, and as a result, several alignment nuclei can be found. Since it is impossible to tell *a priori* which of found nuclei belongs to a true alignment (produced by sequence in the genome that generated a particular read), we compile and save a list of all found nuclei, in order to use them for finding spliced and unsplaced alignments at later stages.

Next, for each nucleus in a list, we search for a possibility of extending alignment without significant gaps and mismatches – at this stage maximum gap size is set to one base pair, i.e. at this step we find extended alignment regions that doesn't contain any introns. We call such alignment intronless. The constructed alignment may contain some unaligned sequences on its flanks: see Figure 2a.

**Figure 2a**. Example of an extended nucleus corresponding to intronless alignment with unaligned flanks that contain potential splice sites (highlighted in bold).

```
          1        11  29262154  29262164  29262174  29262184  29262194  29262204
          nnnnnnnnnnnnnnnn(..)cccagcaccagtacctACCCGTGCCTGTGCCTGTGGGGTCGCCCCCTTGGATCtgtgaaa
          ................(..)................||||||||||||||||| ||||||| |||||||||||.......
          ----------------(..)cccggtatgactccccACCCGTGCCTGTGCCTGGGGGGTCGACCCCTTGGATCacaaagt
         10        10        10        20        30        40        50        60

  29262214  51304551  51304558
          acaaaagcc(..)nnnnnnnnnnnnnnnn
          .........(..)................
          tccggat--(..)----------------
         70        77        77
```

In cases when unaligned tails are long enough (more than 6 bp), we try to find potential splice sites in the vicinity of an alignment boundary (no more than 16 nucleotides away). If a potential splice site cannot be found, but the alignment is good enough, it is still kept as an intronless alignment. In case a splice site is found, an extended alignment is discarded, while its nucleus is used in search for an intron-containing alignment.

Such search is performed in two stages (see Figure 1): First we try to align unaligned flanks with a sequence in the vicinity of an alignment nucleus nucleus. The distance between two alignment blocks shall be no shorter than minimal size of an intron found in a given genome, and no longer than 99th percentile of distribution of intron lengths for that genome. As a result, we get a multiblock alignment (Fig. 2b) with boundaries that don't necessarily correspond to actual splice sites.

**Figure 2b.** Example of a multi-block alignment before attempting to find correct intron/exon boundaries. Nucleotides that can be attributed to more than one block are shown in bold.

```
        1         11  29261504  29261514  29261524  29261534  29261544  29261554
        nnnnnnnnnnnnnnnn(..)cggaactcgtctttgaAGGGCTTCCCCCAGTATGACTCCCCacctggaagagaaagg(..
        ................(..)................|||||||||||| ||||||||||||................(..
        ----------------(..)----------------AGGGCTTCCCCCGGTATGACTCCCC----------------(..
        1          1         1         1         5         15        25        26

29262154  29262163  29262173  29262183  29262193  29262203  29262213  29262531
        )cccagcaccagtacctACCCGTGCCTGTGCCTGTGGGGTCGCCCCCTTGGATCtgtgaaaacaaaagcc(..)actctc
        )................||||||||||||||||| ||||||| |||||||||||................(..)......
        )----------------ACCCGTGCCTGTGCCTGGGGGGTCGACCCCTTGGATC----------------(..)------
      26        26        29        39        49        59        63        63

29262537  29262547  29262557  29262567  51304551  51304557
        gtcactcaccACAAAGTTCCGGATggatctgtggaagatg(..)nnnnnnnnnnnnnnnn
        ..........||||||||||||||................(..)................
        ----------ACAAAGTTCCGGAT----------------(..)----------------
      63        63        73        77        77        77
```

Then, we attempt to bring alignment block boundaries in correspondence with splice sites. The algorithm searches for an optimal solution, taking into account relative weights of different potential splice sites, as well as number of mismatches in a resulting alignment (Fig. 3). If an alignment obtained in such optimization procedure doesn't pass the filtering criteria of quality, it is discarded, and its original nucleus is excluded from further consideration.

**Figure 3**. An example of intron-containing (spliced) multiblock alignment: some nucleotides (shown in bold) were transferred from one block to another in order to correspond the correct intron boundary.

```
1        11  29261504  29261514  29261522  29261532  29261542  29261551
          nnnnnnnnnnnnnnnn(..)cggaactcgtctttga  AGGGCTTCCCCCAGTATGACTCCCCAC]ctggaagagaaagg
          ................(..)................  |||||||||||| |||||||||||||| ..............
          ----------------(..)----------------  AGGGCTTCCCCCGGTATGACTCCCCAC --------------
          1         1         1         1         3         13        23        28

   29261561  29262160  29262170  29262179  29262189  29262199  29262208  29262218
          cc(..)cagcaccagtacctac[CCGTGCCTGTGCCTGTGGGGTCGCCCCCTTGGAT]ctgtgaaaacaaaagc(..)ga
          ..(..)................ ||||||||||||||| ||||||| |||||||||| ................(..)..
          --(..)---------------- CCGTGCCTGTGCCTGGGGGGTCGACCCCTTGGAT ----------------(..)--
         28        28        28        35        45        55        62        62

   29262532  29262542  29262551  29262561  29262569  51304551  51304559
          ctctcgtcactcac[CACAAAGTTCCGGAT  ggatctgtggaagatg(..)nnnnnnnnnnnnnnnn
          .............. |||||||||||||||  ................(..)................
          -------------- CACAAAGTTCCGGAT  ----------------(..)----------------
         62        62        67        77        77        77        77
```

Since search for introns is a very resource-intensive task, such approach – using only some, most promising alignment nuclei, greatly reduces computation time. For instance, for reads of 76 bp length about 38% of all nuclei are excluded from consideration. Such optimization approach, however, has one limitation: its advantage will be lost if reads are longer than average exon length for a given genome. Furthermore, if paired reads are being used, only those alignments that can accommodate both reads from a pair are kept. Thus, as a result of this first stage, we have lists of spliced and unspliced alignments.

### Recreating introns and selecting accurate spliced alignments

Using a set of spliced alignments from the list constructed earlier, we can compile a list of introns and estimate their reliability based on frequencies of their occurrence in the alignments. The list would contain a multitude of variants with unaligned flanks too short (less than 6 bp) to reliably tell whether each of such flanks corresponds to another exon or is a result of sequencing error. (Fig. 4A). However, if alignments of other reads confirm existence of an exon between aligned and unaligned parts of a given read and, at the same time, if unaligned part is similar with a terminus of an adjacent exon, we can, with high degree of conviction, align it to that exon (Fig. 4B, C). This approach lets reliably transfer even short fragments (down to 1 bp) to an adjacent exon. Such procedure is performed on all alignments with short unaligned flanks.

**Figure 4**. Rebuilding intron from short unaligned flanks.

A. An example of initial alignment. There is a short 3-bp unaligned flank at the left, shown in bold.

```
   1        11  16268119  16268129  16268139  16268149  16268159  16268169
       nnnnnnnnnnnnnnnn(..)ttgaatataaaagtatACCTTTCTATCACCACCCTTATTTATTTCTGGTTCTTGAGACAT
       ................(..)................||||||||||||||||||||||||||||||||||||||||||||
       ----------------(..)-------------tcaACCTTTCTATCACCACCCTTATTTATTTCTGGTTCTTGAGACAT
       1         1         1         1         8        18        28        38

16268179  16268189  51304551  51304558
       TTCctgcagatgcaaaaac(..)nnnnnnnnnnnnnnnn
       |||................(..)................
       TTC----------------(..)----------------
      48        51        51        51
```

B. A multiblock (spliced) alignment that supports a potential intron in location shown in Fig. 4A. Canonical splice sites (AC-CT) are present, and block sizes are sufficient to judge this alignment to be correct.

```
       1         11  16267065  16267075  16267083  16267093  16267102  16268121
       nnnnnnnnnnnnnnnn(..)tacttccgtgcttctt  CATTTCTTCTTCAAC]cttgaatgaaagtttg(..)gaatat
       ................(..)................  ||||||||||||||| ................(..)......
       ----------------(..)----------------  CATTTCTTCTTCAAC ----------------(..)------
       1         1         1         1         3        13        16        16

16268127  16268137  16268146  16268156  16268166  16268174  16268184  51304553
       aaaagtatac[CTTTCTATCACCACCCTTATTTATTTCTGGTTCTT  gagacatttcctgcag(..)nnnnnnnnnnnn
       .......... |||||||||||||||||||||||||||||||||||  ................(..)............
       ---------- CTTTCTATCACCACCCTTATTTATTTCTGGTTCTT  ----------------(..)------------
      16        16        25        35        45        51        51        51
```

C. The result of rebuilding an intron. Based on supporting alignment, five nucleotides (three of an unaligned flank from Fig. 4A plus two aligned, all shown in bold) were attributed to a new exon. As a result, the read was completely aligned, and intron position was correctly mapped.

```
        1         11  16267075  16267085  16267095  16267104  16268121  16268129
        nnnnnnnnnnnnnnnn(..)cttcttcatttcttctTCAAC]cttgaatgaaagtttg(..)gaatataaaagtatac[C
        ................(..)................||||| ................(..)................ |
        ----------------(..)----------------TCAAC ----------------(..)---------------- C
        1         1         1         1         5         6         6         6

16268138  16268148  16268158  16268168  16268178  16268188  51304551  51304557
        TTTCTATCACCACCCTTATTTATTTCTGGTTCTTGAGACATTTCctgcagatgcaaaaac(..)nnnnnnnnnnnnnnnn
        ||||||||||||||||||||||||||||||||||||||||||||................(..)................
        TTTCTATCACCACCCTTATTTATTTCTGGTTCTTGAGACATTTC----------------(..)----------------
        7         17        27        37        47        51        51        51
```

The result of this stage is a set of variants of alignments of reads to contigs, taking into account possible splice sites. Based on homology level and lengths of alignments, we can now assign scores to each alignment of a read or a pair of reads, and use these scores to map reads (or pairs) to contigs/chromosomes. I.e., for each read or pair, we find an alignment with maximum score, and discard all alternative alignments with scores below 96% of that maximum. Thus, for each read or pair, we are left with only highest-scoring alignments – usually just one. Such procedure, however, changes the composition of introns, as some of the alignments that supported introns are discarded. So the rebuilding of introns, as discussed above, is performed one more time.

**Testing performance of reads mapping.**

Accuracy of RedsMap and other programs and was tested on an artificially created set of reads with known positions in the human genome sequence. We tested the program on both spliced reads (generated RNA-seq reads) and on unspliced reads (generated genomic reads). In a former case, we took an annotated human chromosome 22 sequence (http://genome.ucsc.edu/, h19 set), extracted from it 1,356 mRNA sequences, and used SimSeq program to simulate Illumina reads with randomly inserted mutations as provided by SimSeq default error profile and with 40x coverage. To estimate effect of reads length on mapping quality, we used three sets of reads, with lengths of 50, 76 and 100 bp. Parameters of each set are shown in Table 1.

**Table 1**. Parameters of test sets: RNA-Seq reads. Potentially spliced highly homologous reads, no insertions/deletions, error profile simulates Illumina sequencer.

| Read length | Number of reads | Number of spliced reads | Parameters of reads |
|---|---|---|---|
| 50 bp | 2,979,624 | 492,743 (16.5%) | insert size = 200 bp, standard deviation = 20 bp, coverage = 40 |
| 76 bp | 1,960,300 | 485,857 (24.8%) | insert size = 200 bp, standard deviation = 20 bp, coverage = 40 |
| 100 bp | 1,489,796 | 469,319 (33.3%) | insert size = 300 bp, standard deviation = 30 bp, |
| | | | coverage = 40 |

For genomic (non-spliced) reads we introduced nucleotide substitutions, short deletions and insertions (up to 4 bp) to the sequence of human chromosome 22 and then used SimSeq program to generated reads (Table 2).

**Table 2**. Parameters of test sets: genomics reads.

| Read length | Number of reads | %mutations | % InDel | Parameters of reads |
|---|---|---|---|---|
| 76bp | 18,363,068 | 0.5 | 0.002 | insert size = 200 bp, standard deviation = 20 bp, coverage = 40 |
| 76bp | 18,363,276 | 1 | 0.02 | insert size = 200 bp, standard deviation = 20 bp, coverage = 40 |
| 76bp | 18,368,502 | 2 | 0.02 | insert size = 200 bp, standard deviation = 20 bp, coverage = 40 |
| 76bp | 18,361,496 | 3 | 0.02 | insert size = 200 bp, standard deviation = 20 bp, coverage = 40 |
| 76bp | 18,365,644 | 4 | 0.02 | insert size = 200 bp, standard deviation = 20 bp, coverage = 40 |
| 76bp | 18,361,920 | 5 | 0.02 | insert size = 200 bp, standard deviation = 20 bp, coverage = 40 |
| 76bp | 18,364,062 | 6 | 0.02 | insert size = 200 bp, standard deviation = 20 bp, coverage = 40 |
| 76bp | 18,369,140 | 7 | 0.02 | insert size = 200 bp, standard deviation = 20 bp, coverage = 40 |
| 76bp | 18,367,384 | 8 | 0.02 | insert size = 200 bp, standard deviation = 20 bp, coverage = 40 |
| 76bp | 18,373,472 | 9 | 0.02 | insert size = 200 bp, standard deviation = 20 bp, coverage = 40 |
| 76bp | 18,371,406 | 10 | 0.02 | insert size = 200 bp, standard deviation = 20 bp, coverage = 40 |

We can see from Table 2 that for genomic reads test we will use 11 sets of non-spliced reads with increasing frequency of mismatches (reflecting potential errors of sequencing). The programs were run on four cores of 3.6-GHz AMD FX 8150 processor, available memory 16 GB.

Results of reads mapping were compared with known reads locations in chromosome sequence.
In cases when we have short unaligned tails the coordinate of alignment extended to the reads' ends abolishing effect of mutations there. We estimate the following measures of program performance [9,10]:

1. Total number of reads which have alignment to genome sequence (that can be correct and not)
2. Total number of reads alignments

3. Number of correct alignments
4. Specificity (Sp) – the ratio of correct alignments to all found alignments
5. Sensitivity (Sn) - the ratio of correct alignments to the number of reads
6. F1 score: 2*(Sn*Sp)/(Sn+Sp)
7. G-measure =sqrt (Sn*Sp)

**Accuracy for potentially spliced (RNASeq reads).**

Results of reads mapping from the test sets by ReadsMap program are presented in Table 3 (a,b,c).

**Table 3. Results of accuracy test of ReadsMap program.**

**A. Reads length 50 bp**

| | Reads | Aligned (Percent) | # Alignments | Correct alignments | Sp | Sn |
|---|---|---|---|---|---|---|
| Unspliced | 2 486 387 | 2486163 (0.99991) | 2600229 | 2482336 | 0.95466 | 0.99837 |
| Spliced | 493 237 | 492977 (0.99947) | 502892 | 489450 | 0.97327 | 0.99232 |
| All | 2 979 624 | 2979140 (0.99984) | 3103121 | 2971786 | 0.95768 | 0.99737 |

**B. Reads length 76 bp**

| | Reads | Aligned (Percent) | # Alignments | Correct alignments | Sp | Sn |
|---|---|---|---|---|---|---|
| Unspliced | 1 473 886 | 1473873 (0.99999) | 1526213 | 1471937 | 0.96444 | 0.99868 |
| Spliced | 486 414 | 486381 (0.99993) | 493335 | 483693 | 0.98046 | 0.99441 |
| All | 1 960 300 | 1960254 (0.99998) | 2019548 | 1955630 | 0.96835 | 0.99762 |

**C. Reads length 100 bp**

| | Reads | Aligned (Percent) | # Alignments | Correct alignments | Sp | Sn |
|---|---|---|---|---|---|---|
| Unspliced | 1 020 477 | 1020444 (0.99997) | 1049899 | 1019278 | 0.97083 | 0.99883 |
| Spliced | 469319 | 469156 (0.99965) | 474509 | 465782 | 0.98161 | 0.99246 |
| All | 1 489 796 | 1489600 (0.99987) | 1524408 | 1485060 | 0.97419 | 0.99682 |

The above results (Table3) show that ReadsMap found correct alignments for more than 99.8% o non-spliced reads and more than 99.4% for spliced reads. Analyzing cases of incorrect alignments we noted that the most often problem was finding spliced alignments for actually non-spliced reads. Such problem can appear due to occurrence of alternative splicing, when a particular gene has several possible isoforms and the found intron (for a considered read) is exist in one of isoform, but this read originally extracted from an exon of alternative isoform (Fig. 5A).

A good example of such situation is alternative isoforms of GAB4 gene: uc002zlw.3

(http://genome.ucsc.edu/cgi-bin/hgGene?hgg_gene=uc002zlw.2) and uc010gqs.1 (http://genome.ucsc.edu/cgi-bin/hgGene?hgg_gene=uc010gqs.1) (Fig. 5 b,c,d).

**Figure 5.** An example of problem with reads mapping due to alternative splicing.

**A.** Alternatively spliced isoforms of GAB4 gene (a and b);  c and d are two possible read alignments; 1,2,4 are exons of isoforms a and b; 3 is the short region with the same nucleotide sequence; arrows are intron sequences.

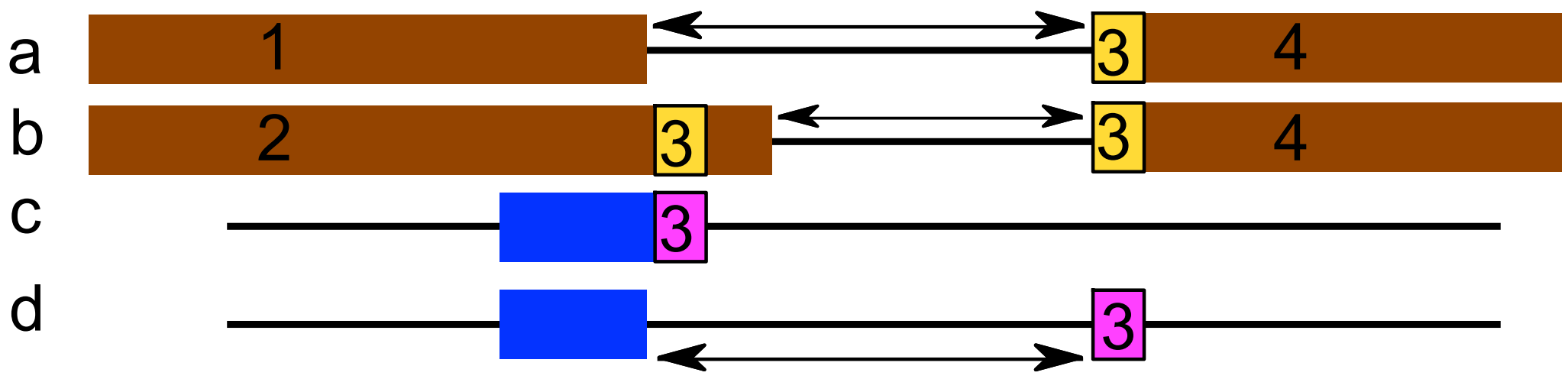

**B.** “Incorrect” mapping variant of spliced alignment. Nucleotides aligned with the wrong exon shown in bold.

```
1        11  17468918  17468928  17468938  17468948  17468958  17468968
          nnnnnnnnnnnnnnnn(..)ccaggtgggcggcacaCAGTGAGACACCGGGGGCTCAGATGTGGGTTCTTGTTCTTGGGG
          ................(..)................||||||||||||||||||||||||||||||||||||||||||||
          ----------------(..)----------------CAGTGAGACACCGGGGGCTCAGATGTGGGTTCTTGTTCTTGGGG
          1         1         1         1         5        15        25        35

   17468978  17468988  17468998  17469007  17469017  17472747  17472757  17472766
          GAGGTGCTGATGGGAGCAGCTGGGCTCAG]ctggagaagagcagag(..)gtgtgtgctcccatac[CTGtgctttcctc
          ||||||||||||||||||||||||||||| ................(..)................ |||..........
          GAGGTGCTGATGGGAGCAGCTGGGCTCAG ----------------(..)---------------- CTG----------
         45        55        65        74        74        74        74        77
```

**C.** Correct spliced alignment including actual intron sequence between exons 17468850...17469006 and 17472763...17473066 of chr22 uc002zlw.2 gene.

```
          1        11  17469017  17469027  17469035  17469045  17469055  17469064
          nnnnnnnnnnnnnnnn(..)gcagaggccgtgactg  GCTGAGGAAATGTTTCCCAGGAAGC]ctgtgaatgaaacagg
          ................(..)................  ||||||||||||||||||||||||| ................
          ----------------(..)----------------  GCTGAGGAAATGTTTCCCAGGAAGC ----------------
          1         1         1         1         3        13        23        26

   17472747  17472753  17472763  17472772  17472782  17472792  17472802  17472812
          (..)gtgtgtgctcccatac[CTGTGCTTTCCTCCTGCCTGAAGCCACAGATCTGACAGATGCTCTGGACCC actcat
          (..)................ ||||||||||||||||||||||||||||||||||||||||||||||||||| ......
          (..)---------------- CTGTGCTTTCCTCCTGCCTGAAGCCACAGATCTGACAGATGCTCTGGACCC ------
         26        26        26        35        45        55        65        75
```

**D.** Correct unspliced alignment when the reads included in exon (17468850...17469057) of uc010gqs.1 gene. The nucleotides that mapped to alternative exon in the above spliced alignment shown in bold.

```
         1         11  17468918  17468928  17468936  17468946  17468956  17468966
         nnnnnnnnnnnnnnnn(..)ccaggtgggcggcaca  CAGTGAGACACCGGGGGCTCAGATGTGGGTTCTTGTTCTTGG
         ................(..)................  ||||||||||||||||||||||||||||||||||||||||||
         ----------------(..)----------------  CAGTGAGACACCGGGGGCTCAGATGTGGGTTCTTGTTCTTGG
         1         1         1         1         3        13        23        33

 17468976  17468986  17468996  17469006  17469014  17469024  51304555  51304565
         GGGAGGTGCTGATGGGAGCAGCTGGGCTCAGCTG  gagaagagcagaggcc(..)nnnnnnnnnnnnnnnn
         ||||||||||||||||||||||||||||||||||  ................(..)................
         GGGAGGTGCTGATGGGAGCAGCTGGGCTCAGCTG  ----------------(..)----------------
        43        53        63        73        77        77        77        77
```

Mapping variants (c and d) represent equally good read ‘s alignments. Having in mind potential usage of the program for identification of gene structure and its alternative isoforms we select the spliced variant if we consider just one mapping for a read.

The following tables include results of testing several popular reads mapping programs.

**TopHat** (http://ccb.jhu.edu/software/tophat/index.shtml) is one of the most accurate programs for aligning RNASeq reads to genomic sequences [11].

**Table 4.** Results of accuracy test of **TopHat** v 2.01 program.

**A. Reads length 50 bp**

| | Reads | Aligned (Percent) | # Alignments | Correct alignments | Sp | Sn |
|---|---|---|---|---|---|---|
| Unspliced | 2 486 387 | 2482498 (0.99844) | 2665115 | 2481320 | 0.93104 | 0.99796 |
| Spliced | 493 237 | 487403 (0.98817) | 1059827 | 480087 | 0.88179 | 0.97334 |
| All | 2 979 624 | 2969901 (0.99674) | 3209564 | 2961407 | 0.92268 | 0.92268 |

**B. Reads length 76 bp**

| | Reads | Aligned (Percent) | # Alignments | Correct alignments | Sp | Sn |
|---|---|---|---|---|---|---|
| Unspliced | 1 473 886 | 1461911 (0.99188) | 1519845 | 1461782 | 0.96180 | 0.99179 |
| Spliced | 486 414 | 476074 (0.97874) | 515719 | 471919 | 0.91507 | 0.97020 |
| All | 1 960 300 | 1937985 (0.98862) | 2035564 | 1933701 | 0.94996 | 0.98643 |

**C. Reads length 100 bp**

| | Reads | Aligned (Percent) | # Alignments | Correct alignments | Sp | Sn |
|---|---|---|---|---|---|---|
| Unspliced | 1 020 477 | 950942 (0.93186) | 982077 | 1433124 | 0.96803 | 0.93161 |
| Spliced | 469319 | 420868 (0.89676) | 451047 | 1433124 | 0.92752 | 0.89141 |
| All | 1 489 796 | 1371810 (0.92080) | 1433124 | 1433124 | 0.95528 | 0.91894 |

We can see that while the accuracy of mapping unspliced reads by **TopHat** is practically the same as for **ReadsMap** program, the performance for unspliced reads is slightly worse.

**STAR** (https://code.google.com/p/rna-star/) program is known as ultrafast universal RNA-seq aligner [12].

**Table 5.** Results of accuracy test of **STAR** v 2.4.0 program.

### A. Reads length 50 bp

| | Reads | Aligned (Percent) | # Alignments | Correct alignments | Sp | Sn |
|---|---|---|---|---|---|---|
| Unspliced | 2 486 387 | 2483628 (0.99889) | 2625514 | 2482625 | 0.94558 | 0.99849 |
| Spliced | 493 237 | 491281 (0.99603) | 517341 | 319885 | 0.61833 | 0.64854 |
| All | 2 979 624 | 2974909 (0.99842) | 3142855 | 2802510 | 0.89171 | 0.94056 |

### B. Reads length 76 bp

| | Reads | Aligned (Percent) | # Alignments | Correct alignments | Sp | Sn |
|---|---|---|---|---|---|---|
| Unspliced | 1 473 886 | 1473804 (0.99994) | 1532735 | 1473598 | 0.96142 | 0.99980 |
| Spliced | 486 414 | 485988 (0.99912) | 369229 | 369229 | 0.73010 | 0.75908 |
| All | 1 960 300 | 1959792 (0.99974) | 1842827 | 1842827 | 0.90403 | 0.94007 |

### C. Reads length 100 bp

| | Reads | Aligned (Percent) | # Alignments | Correct alignments | Sp | Sn |
|---|---|---|---|---|---|---|
| Unspliced | 1 020 477 | 1020441 (0.99996) | 1057264 | 1019945 | 0.96470 | 0.99948 |
| Spliced | 469319 | 468786 (0.99886) | 486121 | 366995 | 0.75495 | 0.78197 |
| All | 1 489 796 | 1489227 (0.99962) | 1543385 | 1386940 | 0.89864 | 0.93096 |

**PASS** [7,8] (http://pass.cribi.unipd.it/cgi-bin/pass.pl) can map non-spliced and spliced reads. It was run with the following options: ./pass -cpu 4 -p 1111110111111 -b -flc 1 -fid 90 -sam -phred64 for analysis of unspliced reads and ./pass -cpu 4 -p 1111110111111 -b -flc 1 -fid 90 -sam -phred64 -spliced rna -percent_tolerance 30 -fle 10 for spliced reads. Pass mapping results are presented in Table 6.

**Table 6.** Results of accuracy test of **PASS** program.

**A. Reads length 50 bp**

| | Reads | Aligned (Percent) | # Alignments | Correct alignments | Sp | Sn |
|---|---|---|---|---|---|---|
| Unspliced | 2 486 387 | 2 486 369 (0.99999) | 2 630 353 | 2 478 444 | 0.94225 | 0.99681 |
| Spliced | 493 237 | 420 143 (0.85181) | 434 378 | 235 700 | 0.54261 | 0.47786 |
| All | 2 979 624 | 2 906 512 (0.97546) | 3 064 731 | 2 714 144 | 0.88561 | 0.91090 |

**B. Reads length 76 bp**

| | Reads | Aligned (Percent) | # Alignments | Correct alignments | Sp | Sn |
|---|---|---|---|---|---|---|
| Unspliced | 1 473 886 | 1 473 886 (1.00000) | 1 529 970 | 1 471 884 | 0.96203 | 0.99864 |
| Spliced | 486 414 | 459 310 (0.94428) | 471 255 | 282 435 | 0.59933 | 0.58065 |
| All | 1 960 300 | 1 933 196 (0.98617) | 2 001 225 | 1 754 319 | 0.87662 | 0.89492 |

**C. Reads length 100 bp**

| | Reads | Aligned (Percent) | # Alignments | Correct alignments | Sp | Sn |
|---|---|---|---|---|---|---|
| Unspliced | 1 020 477 | 1 020 477 (1.00000) | 1 051 803 | 1 019 096 | 0.96890 | 0.99865 |
| Spliced | 469 319 | 450 956 (0.96087) | 460 505 | 270 255 | 0.58687 | 0.57585 |
| All | 1 489 796 | 1 471 433 (0.98767) | 1 512 308 | 1 289 351 | 0.85257 | 0.86545 |

Most of errors of PASS program are due to poor finding short alignments blocks (less than 7 bp) at the ends of introns. Therefore it would be interesting to estimate accuracy of finding the main parts of alignments. For this task we selected the longest block of alignment (not interrupted by intron) and estimate the accuracy of correct finding 5 middle positions of such block (Table 6).

**Table 6.** Results of accuracy test of finding longest block of alignments by **PASS** program.

**A. Reads length 50 bp**

| | Reads | Aligned (Percent) | # Alignments | Correct Block alignments | Sn |
|---|---|---|---|---|---|
| Unspliced | 2 486 387 | 2 486 369 (0.99999) | 2 630 353 | 2 478 444 | 0.99681 |
| Spliced | 493 237 | 420 143 (0.85181) | 434 378 | 415 086 | 0.84155 |
| All | 2 979 624 | 2 906 512 (0.97546) | 3 064 731 | 2 893 530 | 0.97111 |

## B. Reads length 76 bp

| | Reads | Aligned (Percent) | # Alignments | Correct Block alignments | Sn |
|---|---|---|---|---|---|
| Unspliced | 1 473 886 | 1 473 886 (1.00000) | 1 529 970 | 1 471 884 | 0.99864 |
| Spliced | 486 414 | 459 310 (0.94428) | 471 255 | 452 608 | 0.93050 |
| All | 1 960 300 | 1 933 196 (0.98617) | 2 001 225 | 1 924 492 | 0.98173 |

## C. Reads length 100 bp

| | Reads | Aligned (Percent) | # Alignments | Correct Block alignments | Sn |
|---|---|---|---|---|---|
| Unspliced | 1 020 477 | 1 020 477 (1.00000) | 1 051 803 | 1 019 096 | 0.99865 |
| Spliced | 469 319 | 450 956 (0.96087) | 460 505 | 439 890 | 0.93729 |
| All | 1 489 796 | 1 471 433 (0.98767) | 1 512 308 | 1 458 986 | 0.97932 |

We can see that in most mappings Pass program is pretty effective to find correctly the longest block of alignment, while the results are depending on the length of reads. For reads with 50 bp length the accuracy was just 87% and it increased to 93% for longer reads. Note that the errors due to imperfect alignment of short sequences separated by long intron from the other part of alignment are observed in all tested programs, but the ReadsMap has the least number of them.

Sensitivity (Sn) and specificity (Sp) measures allow evaluating a program performance, while their values alone may be highly misleading. Therefore, we computed F1-score and G-measure that account both of them providing a single value of accuracy, which is useful to rank the tested programs performance (Table 7).

**Table 7. The summary table of accuracy spliced reads mapping programs.**

| | 50 bp | | 76 bp | | 100 bp | |
|---|---|---|---|---|---|---|
| | Sn<br>Sp | F1-Score<br>G-Measure | Sn<br>Sp | F1-Score<br>G-Measure | Sn<br>Sp | F1-Score<br>G-Measure |
| ReadsMap | **0.99737**<br>**0.95768** | **0.97712**<br>**0.97732** | **0.99762**<br>**0.96835** | 0.98276<br>0.98288 | 0.99682<br>0.97419 | 0.98537<br>0.98544 |
| TopHat | 0.99389<br>0.92268 | 0.95696<br>0.95762 | 0.98643<br>0.94996 | 0.96785<br>0.96802 | 0.91894<br>0.95528 | 0.93676<br>0.93693 |
| Star | 0.94056<br>0.89171 | 0.91548<br>0.91581 | 0.94007<br>0.90403 | 0.92170<br>0.92187 | 0.93096<br>0.89864 | 0.91451<br>0.91466 |
| Pass | 0.91547<br>0.89005 | 0.90258<br>0.90267 | 0.90603<br>0.88750 | 0.89667<br>0.89672 | 0.87765<br>0.86458 | 0.87107<br>0.87109 |

These results is in agreement with the published accuracy estimations for TopHat, Star and PASS [8,11,12].

The current version of ReadsMap program has superior accuracy after correction of some errors found in its earlier versions. At the same time execution of ReadsMap require more computational resources than for the other tested programs (Table 8).

**Table 8**. Average time of alignment for 1000 sequnces (sec).

| | |
|---|---|
| ReadsMap | 5,17 |
| TopHat | 0,47 |
| Star | 0,09 |
| Pass | 0,91 |

## Mapping genomic (DNAseq) reads.

Alignment of genomic reads with the reference genome is easier than RNASeq reads, while its high accuracy is very important for identification of SNP. Bowtie and BWA are the most popular programs for this task. In Tables 9-11 we provide various measures of accuracy and execution time for ReadsMap, Bowtie and BWA.

**Table 9**. Mapping results for **ReadsMap** program.

| Length, bp | Number | Aligned (Percent) | #Alignments | Correct | Sp | Sn | F1-Score | G-Measure | Time, sec |
|---|---|---|---|---|---|---|---|---|---|
| 100 | 1 020 477 | 1020477 (1.00000) | 1020477 (1.0000) | 1020424 | 0.96978 | 0.99995 | 0.98463 | 0.98475 | 548.03 |
| 76 | 1 473 886 | 1473886 (1.0000) | 1530231 | 1473831 | 0.96314 | 0.99996 | 0.98120 | 0.98138 | 452.56 |
| 50 | 2 486 387 | 2486336 (0.99998) | 2711089 | 2485726 | 0.91687 | 0.99973 | 0.95651 | 0.95740 | 980.28 |

We can see that the number of found correct alignments is close to 100%.

The results of applying Bowtie program (http://bowtie-bio.sourceforge.net/index.shtml) to alignment of the same set of reads are presented in Table. 10. The program was executed with "–sensitive" option.

**Table 10**. Mapping results for **Bowtie** program.

| Length, bp | Number | Aligned (Percent) | #Alignments | Correct | Sp | Sn | F1-Score | G-Measure | Time,sec* |
|---|---|---|---|---|---|---|---|---|---|
| 100 | 1 020 477 | 1020244 (0.99977) | 1020244 | 1009052 | 0.98903 | 0,98880 | 0.98891 | 0.98891 | 179.26 |
| 76 | 1 473 886 | 1473158 (0.99951) | 1473158 | 1454956 | 0.98764 | 0,98716 | 0.98740 | 0.98740 | 156.86 |
| 50 | 2 486 387 | 2477883 (0.99658) | 2477883 | 2434210 | 0.98237 | 0.97901 | 0.98069 | 0.98069 | 158.74 |

* Time without indexing of reference genome; indexing time ~60 sec.

Having slightly higher specificity Bowtie demonstrate slightly lower sensitivity comparing with ReadsMap results.

Table 11 includes the results of applying BWA program (v. 0.7.12: http://bio-bwa.sourceforge.net). For reads alignment we used its default parameters. Note that the accuracy of BWA significantly less depends on length of reads and this software shows maximal specificity among the tested programs, while as in the case with Bowtie, ~ for 1% of reads it has not found their correct alignments.

**Table 11**. Mapping results for **BWA** program.

| Length, bp | Number reads | Aligned (Percent) | #Alignments | Correct | Sp | Sn | F1-Score | G-Measure | Time,sec* |
|---|---|---|---|---|---|---|---|---|---|
| 100 | 1 020 477 | 1019620 (0.99916) | 1019620 | 1009030 | 0,98961 | 0,98878 | 0,98919 | 0,98919 | 347.29 |
| 76 | 1 473 886 | 1473210 (0.99954) | 1473210 | 1455150 | 0,98774 | 0,98729 | 0,98751 | 0,98751 | 322.25 |
| 50 | 2 486 387 | 2485349 (0.99958) | 2485349 | 2441869 | 0,98251 | 0,98210 | 0,98230 | 0,98230 | 297.23 |

* Time without indexing of reference genome; indexing time ~65 sec.

In the following three tables we presented results of mapping DNASeq reads with increasing number of errors or possible SNP (from 0.5 to 10%).

**Table 12**. Mapping results for reads with high mutation rate by ReadsMap.

| % mutations | Number reads | Aligned (Percent) | #Alignments | Correct | Sp | Sn | F1-Score | G-Measure | Time, sec (real/ user) |
|---|---|---|---|---|---|---|---|---|---|
| 0,5 | 18 363 068 | 18362950 (0.99999) | 19254664 | 18337682 | 0.95601 | 0.99862 | 0,97685 | 0.97708 | 19808.01/ 78731.24 |
| 1 | 18 363 276 | 18234690 (0.99300) | 19062577 | 17002934 | 0.89195 | 0.92592 | 0,90862 | 0.90878 | 18114.58/ 72297.28 |
| 2 | 18 368 502 | 18242336 (0.99313) | 19072911 | 14514207 | 0.76099 | 0.79017 | 0,77531 | 0.77544 | 12882.56/ 51446.78 |
| 3 | 18 361 496 | 18217040 (0.99213) | 19050118 | 15168519 | 0.79624 | 0.82610 | 0,81090 | 0.81104 | 10426.53/ 41601.00 |
| 4 | 18 365 644 | 18150884 (0.98831) | 18977522 | 12308958 | 0.64861 | 0.67022 | 0,65924 | 0.65932 | 8597.43/ 34269.59 |
| 5 | 18 361 920 | 17987030 (0.97958) | 18808113 | 12441751 | 0.66151 | 0.67758 | 0,66945 | 0.66950 | 7194.63/ 28647.59 |
| 6 | 18 364 062 | 17674114 (0.96243) | 18489866 | 11740121 | 0.63495 | 0.63930 | 0,63712 | 0.63712 | 5919.31/ 23539.51 |
| 7 | 18 369 140 | 17151896 (0.93373) | 17936586 | 11161720 | 0.62229 | 0.60763 | 0,61487 | 0.61492 | 5120.06/ 20211.45 |
| 8 | 18 367 384 | 16372418 (0.89139) | 17133257 | 10173917 | 0.59381 | 0.55391 | 0,57317 | 0.57351 | 4882.91/ 19287.23 |
| 9 | 18 373 472 | 15337898 (0.83478) | 16055421 | 9380526 | 0.58426 | 0.51055 | 0,54492 | 0.54616 | 4039.63/ 15972.67 |
| 10 | 18 371 406 | 14019110 (0.76309) | 14676629 | 8014959 | 0.54610 | 0.43627 | 0,48505 | 0.48811 | 3383.01/ 13345.65 |

**Table 13**. Mapping results for reads with high mutation rate by BWA

| % mutations | Number Reads | Aligned (Percent) | #Alignments | Correct | Sp | Sn | F1-Score | G-Measure | Time,sec (real/ user) |
|---|---|---|---|---|---|---|---|---|---|
| 0,5 | 18 363 068 | 18361880 (0.99994) | 18361930 | 17850991 | 0.97217 | 0.97211 | 0,97214 | 0,97214 | 554.54/ 2144.31 |
| 1 | 18 363 276 | 18356434 (0.99963) | 18356584 | 16491998 | 0.89842 | 0.89810 | 0,89826 | 0,89826 | 568.89/ 2274.84 |
| 2 | 18 368 502 | 18319060 (0.99731) | 18319621 | 13913303 | 0.75948 | 0.75745 | 0,75846 | 0,75846 | 698.14/ 2794.73 |
| 3 | 18 361 496 | 18183443 (0.99030) | 18184540 | 14278480 | 0.78520 | 0.77763 | 0,78141 | 0,78140 | 800.88/ 3203.97 |
| 4 | 18 365 644 | 17925841 (0.97605) | 17927588 | 11270353 | 0.62866 | 0.61367 | 0,62112 | 0,62107 | 848.23/ 3394.57 |
| 5 | 18 361 920 | 17493416 (0.95270) | 17495931 | 11016734 | 0.62967 | 0.59998 | 0,61465 | 0,61447 | 906.29/ 3627.16 |
| 6 | 18 364 062 | 16884555 (0.91943) | 16887439 | 9984060 | 0.59121 | 0.54367 | 0,56694 | 0,56644 | 932.19/ 3726.93 |
| 7 | 18 369 140 | 16075768 (0.87515) | 16078915 | 9095045 | 0.56565 | 0.49513 | 0,52922 | 0,52805 | 843.29/ 3376.47 |
| 8 | 18 367 384 | 15102771 (0.82226) | 15106106 | 7934956 | 0.52528 | 0.43201 | 0,47637 | 0,47410 | 808.49/ 3242.94 |
| 9 | 18 373 472 | 13978748 (0.76081) | 13981898 | 7015874 | 0.50178 | 0.38185 | 0,43773 | 0,43368 | 770.71/ 3091.98 |
| 10 | 18 371 406 | 12718462 (0.69230) | 12721240 | 5778036 | 0.45420 | 0.31451 | 0,39726 | 0,38666 | 728.42/ 2923.00 |

**Table 14**. Mapping results for reads with high error rate by Bowtie

| % murations | Number reads | Aligned (Percent) | #Alignments | Good | Sp | Sn | F1-Score | G-Measure | Time,сек |
|---|---|---|---|---|---|---|---|---|---|
| 0,5 | 18 363 068 | 18362950 (0,99999) | 18258558 | 1776980 3 | 0.97323 | 0.96769 | 0.97045 | 0.97046 | 2449.11 |
| 1 | 18 363 276 | 18234432 (0,99298) | 18054372 | 1627654 4 | 0.88636 | 0.88636 | 0.88636 | 0.88636 | 2569.75 |
| 2 | 18 368 502 | 18242134 (0,99312) | 17355459 | 1327252 8 | 0.76475 | 0.72257 | 0.74306 | 0.74336 | 2543.34 |
| 3 | 18 361 496 | 18216744 (0,99212) | 16230075 | 1295174 3 | 0.79801 | 0.70538 | 0.74884 | 0.75027 | 2446.93 |
| 4 | 18 365 644 | 18150184 (0,98827) | 9582292 | 9582292 | 0.64740 | 0.52175 | 0.57782 | 0.58119 | 2277.48 |
| 5 | 18 361 920 | 17985918 (0,97952) | 13129087 | 8662673 | 0.65981 | 0.47177 | 0.55017 | 0.55792 | 2102.27 |
| 6 | 18 364 062 | 17672136 (0,96232) | 11373687 | 7194946 | 0.63260 | 0.39179 | 0.48389 | 0.49784 | 1854.86 |
| 7 | 18 369 140 | 17148746 (0,93356) | 9596446 | 5946099 | 0.61961 | 0.32370 | 0.42524 | 0.44785 | 1701.03 |
| 8 | 18 367 384 | 16367454 (0,89112) | 7885925 | 4663468 | 0.59137 | 0.25390 | 0.35527 | 0.38749 | 1528.09 |
| 9 | 18 373 472 | 15330374 (0,83438) | 6337211 | 3684500 | 0.58141 | 0.20053 | 0.29821 | 0.34145 | 1350.17 |
| 10 | 18 371 406 | 14010072 (0,76260) | 4935904 | 2681500 | 0.54326 | 0.14596 | 0.23010 | 0.28159 | 1199.94 |

From Tables 12-14 we can see that with increasing number of (SNP/errors) BWA and Bowtie programs demonstrate significantly bigger decreasing of mapping accuracy than ReadsMap. At the same time BWA and Bowtie required much less computer resources.

**Conclusion**

Summing the results of testing the accuracy of reads mapping programs we can conclude that for non-spliced reads (DNAseq reads) with small number of errors all of them demonstrate very high accuracy (F1-score ~ 98-99%, Tables 9 - 11) where ReadsMap shows maximal sensitivity among other programs while it is less specific. This effect is especially observed for shorter reads that can produce several alignments (observed in deferent genome locations) with approximately the same quality. In cases when we need to have a high specificity, ReadsMap can achieve it by filtering out alignments with smaller score than the best one (while it will slightly reduce sensitivity).

As the most significant result of our approach we consider a very good accuracy of ReadsMap program in mapping spliced reads that crossing intron sequences in the genome sequence. Such mappings provide accurate intron positions that is important in prediction gene structure and can be used helping to resolve a notoriously difficult problem of identification of alternatively spliced gene isoforms. From the Table 7 we can see that ReadsMap demonstrates superior results comparing with the other reads mapping programs. It is achieved mostly by better placement of very short parts of reads separated by long introns with accounting information from mapping other reads containing the same intron inserted between bigger blocks. However, ReadsMap also have better accuracy when we consider the accuracy of correct alignment only for the biggest part of reads (uninterrupted by intron) for STAR и PASS programs. These results show that hash-based algorithms while require more computational resources can provide more accurate results in difficult cases of spliced reads alignment. Currently ReadsMap program is used along (i.e. [13] ) and as an optional part of Fgenesh++ pipeline [14] that has been used in annotation of many eukaryotic genomes). If RNASeq reads are available with the genomic sequence it significantly improves quality of gene prediction and provides possibility using ReadsMap results to annotate alternatively spliced gene isoforms.